\documentclass[]{article}

\usepackage{lmodern}
\usepackage[english]{babel}
\usepackage{microtype}
\usepackage{tabularx,booktabs}
\usepackage{xspace}
\usepackage{textcomp}
\usepackage{listings}
\usepackage{fancyvrb}
\usepackage{mathtools}
\usepackage{graphicx}
\usepackage{subfigure}
\usepackage{booktabs}
\graphicspath{{./figures/}}
\usepackage{hyperref}
\usepackage{url}

\title{An Object-Oriented Framework for Designing Reusable and Maintainable DEVS Models using Design Patterns}
\author{Maamar El Amine HAMRI\\
 MoFED, \url{https://www.lis-lab.fr/mofed/}\\
Aix-Marseille Université, CNRS, Université de Toulon,\\
 LIS UMR 7020, Marseille, France\\ }

\begin{document}

\maketitle

\begin{abstract}
Design patterns are well practices to share software development experiences. These patterns allow enhancing reusability, readability and maintainability of architecture and code of software applications. As simulation applies computerized models to produce traces in order to obtain results and conclusions, designers of simulation explored design patterns to make the simulation code more reusable, more readable and easy to maintain, in addition to design complex software oriented simulation modeling.

In DEVS (Discrete Event System specification), the designers have successfully designed simulations, frameworks, tools, etc. However, some issues remain still open and should be explored like how a piece of code that implements a set of states, events and transitions may be reused to design a new DEVS model? How may a DEVS model be extended to a new formalism? Etc.

In this paper, we address these issues and we propose a set of patterns that may serve as guidelines to designers of DEVS models and its extensions and may contribute to the design of an operational simulation framework. These patterns are inspired partly by the available designs of DEVS community and software engineering developers.   

\end{abstract}

\section{Introduction}
\label{intro}
The use of Modeling and Simulation (M\&S) increases more and more in industry and teaching. In order to give to this discipline strong foundations, scientists and researchers defined and still define formalisms, methods, tools and theories. One of the famous and popular theories in this discipline is that proposed by Zeigler called Theory of Modeling and Simulation~\cite{Zeigler2018} and on which many applications were developed \cite{Wainer2016}. The Discrete EVent System specification (DEVS) formalism, that is the basis of this theory, allows the separation of the modeling requirements from the simulation. In fact, the simulation algorithm is reused  for making behaviors of different models. However, models should be redesigned each time the user requirements evolve using a rigorous syntax and a concise operational semantics. 

Another interesting feature of DEVS is its ability to propose a general framework to design systems in different behavioral paradigms (continuous and discrete systems). Thanks to the expressiveness of DEVS, this framework may be customized allowing the definition of extensions to solve specific problems at conceptual level.  For example, DEV\&DESS is an extension of DEVS where continuous and discrete event models may be designed and accurate simulations may be conducted in a unique framework \cite{Zeigler2006}.

Often, in a DEVS extension, the modeling still remains independent from the simulation core. Elements employed in model are those extended from the classical DEVS: state, event, atomic model, coupled model, etc. but with a new operational semantics: how a transition is fired, how time elapses, etc. The new simulation core is built with the same building rules of DEVS: the root-coordinator to manage time, for each coupled model its dedicated coordinator to dispatch messages and for each atomic model its basic simulator in order to make behaviors \cite{Zeigler2000}. However, some modifications should be conducted in the new simulation core to answer the requirements of the extension. For example, in DynamicStructure-DEVS (DS-DEVS) \cite{Hu2005}, the simulation core should update its structure each time the model under simulation changes structure. 

Reusability of DEVS simulations has been widely discussed at the conceptual level. However, at design level, reusability of DEVS simulations still remain an open issue of research. Reusability and composability are highly desirable goals in design; which remain difficult to achieve because they require that components work under a range of possible requirements and that can be validated under a range of possible functional and logic requirements \cite{Spiegel2005}. In addition to facilitate the achievement of such goals, the simulation code should be opened for technical requirements of object composability and reusability.

On the other hand, all software tools employ the concept of reuse by providing the design and implementation of basic models (code) then reuse them as black box in order to design new models. The Object Oriented Paradigm (OOP) which many programming languages are built, provides the reuse of runnable models (code) by inheritance and composition of objects. Inheritance consists of subtyping an object in order to add new attributes and methods to it and composition consists of gathering objects to make a new object \cite{Fowler1999}.

Note that DEVS and its extensions reuse models (coupled and atomic) by using composition and inheritance to design runnable models while staying in the same framework (DEVS or an extension). Nevertheless, reusing existing models to design other models in an DEVS extension is not common due to the weakness of commonly used software design techniques. In addition, DEVS does not propose ways to reuse behaviors from atomic models to design new ones and to update them. All atomic models are often designed from scratch. In order to overcome this lack, we propose an extensible software framework for DEVS in which models may be reused in two different ways: 
\begin{enumerate}
	\item reuse models as white boxes instead of black ones in order to update the behavior of atomic models to design new models;
	\item reuse models to define new ones in an extensible framework, ie., designed models in DEVS may be specialized to design new models of the DEVS extension.
\end{enumerate}

Note that software engineering techniques should be emphasized in the M\&S cycle. According to \cite{Robinson2006}, a group of researchers of United Kingdom operational research society simulation workshop identified specific areas for discussion. One of them is the use of software engineering techniques to conduct conceptual modeling and simulation. The survey of \cite{Woodcock2009} showed that techniques like design by contract based on OOP, Object Constraint Language (OCL), assertions, etc. contribute efficiently to the success of software development. On the other hand, the use of design patterns make the design of programs more flexible, modular, reusable and understable \cite{Ampatzoglou2013} \cite{Khwaja2016} \cite{Mayvan2017} \cite{Khomh2018} .

In order to make DEVS runnable models flexible, reusable and maintainable, we explore these different techniques of design in OOP. According to software engineering, reusability of code is the likelihood that a segment of source code can be used again to add new functionalities with a slight or no modification \cite{Fowler1999}\cite{Wikipedia2014}. This criterion joins perfectly our requirement for which we attempt to make simulation models more reusable and easy to maintain.

\section{Motivation}
The DEVS community made and is making a serious effort for standardizing DEVS model representation by allowing specification of DEVS models independent from programming languages \cite{Wainer2010}. This allows re-use of models into different simulation frameworks. However our initiative focuses on proposing designs of DEVS in OOP, so close to the programming languages, in order to have executable models wholly or partially reusable from DEVS to DEVS or to any extension. This starts by separating the structure of a DEVS model (ports ensuring encapsulation) from its behavior (functions ensuring the evolution of state variables and computing outputs) in order to make code maintainable for further re-uses.

\begin{figure}[h]
	{
		\centering
		{\includegraphics[trim= 0mm 20mm 10mm 5mm, clip, height =6cm, width= 4cm]{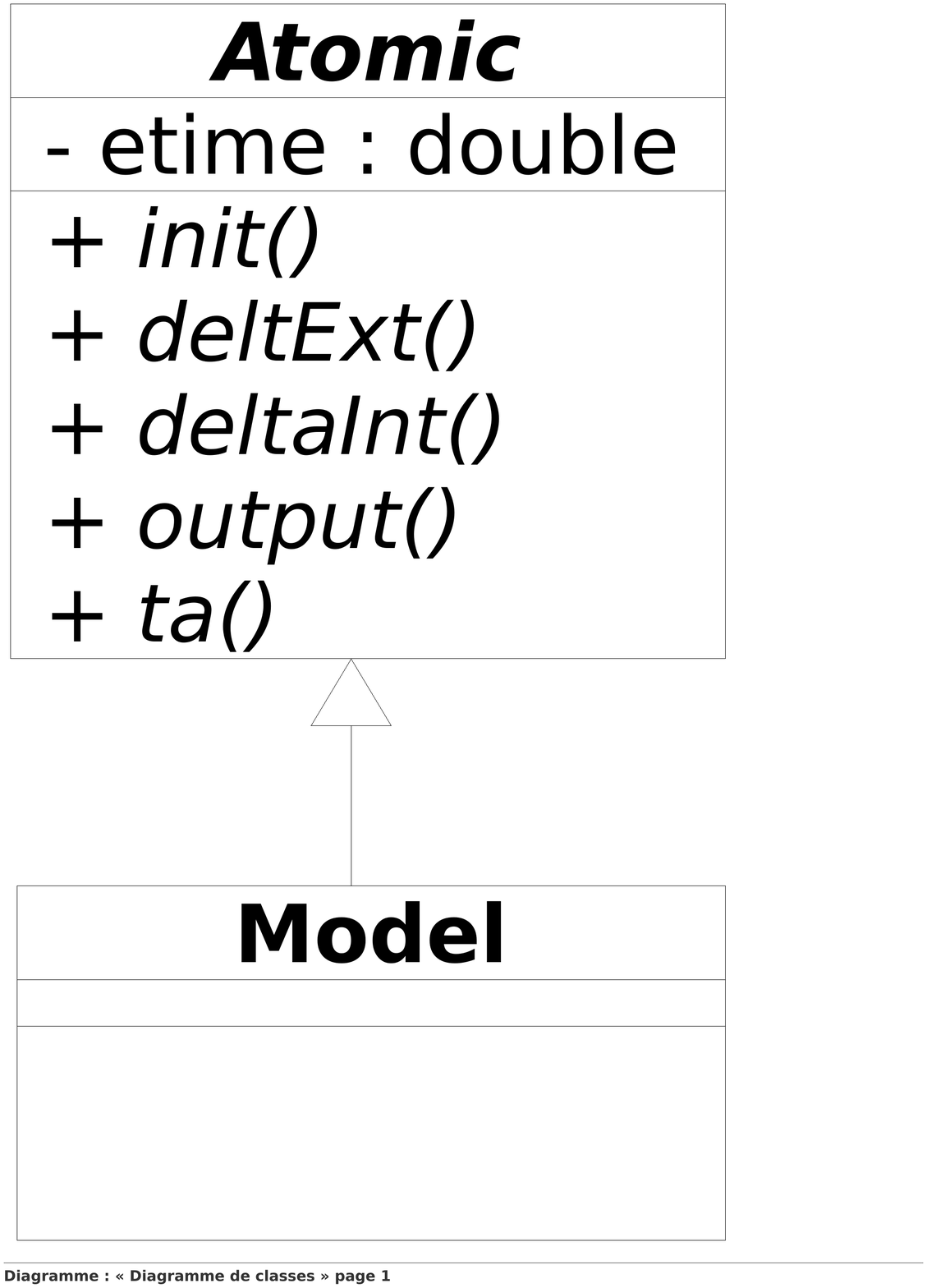}}
		\caption{Basic design of DEVS atomic models.}
		\label{fig:basicdesign}
	}
\end{figure}

Let us consider the class {\tt Model} designing  DEVS atomic models shown in Figure~\ref{fig:basicdesign}.

The use of the class {\tt atomic} as a basis of design of more specialized DEVS, will oblige the subclasses to behave like the basis class.  Such a design can not support designing DEVS extensions as DS-DEVS in which the behavior of atomic models evolves in runtime. In fact, this basic design which consists of only one class does not provide any flexibility to be updated in runtime. On the other hand, the subclasses inherited from class {\tt atomic} are reused as they are. For example, for a given model, if its default state changes, the designer must  override the method {\tt init()} in a new subclass of the considerate class. This must be for the other methods, any new behavior to add into a given class should reuse wholly the behavior of designed methods or overwrite them. Unfortunately, we can not reuse such behaviors due to the atomicity of the expressed behavior.

In OOP, the inheritance, a fundamental principle, allows reusing of code from designed classes. However, it can make  maintainability and design difficult \cite{Barros2015}. For that reason, designers who use OOP, privilege composition and delegation instead of inheritance. In fact, inheritance induces a strong coupling between classes, so reduces maintainability and consequently limit reusability. Stein~\cite{Stein1987} noted that delegation and inheritance are duals and can be used for definition and sharing in software. Consequently, the use of delegation in the proposed framework will contribute efficiently in reusing and maintaining simulation code, without no loss of the power of inheritance.

On our side, in \cite{Hamri2010}, we discussed the advantages of using design patterns by the DEVS community in developing simulations. Furthermore, we enhanced the state patterns to objectify events in addition to states of object  \cite{Hamri2013}\cite{Hamri2014}. These ideas and designs will constitutes the core of our framework in which new designs and implementation of DEVS models may be made and different from those existing.

\section{State of the art}

The reusability and maintainability of DEVS simulations were pointed out by many scientists and engineers. They proposed plug-ins, frameworks, tools and guidelines to enhance the process of developing DEVS simulations.  A list of most useful and popular DEVS simulators may be found in \cite{Franceschini2014} \cite{Van2017} \cite{Goldstein2018} where comparison studies were conducted among cited simulators according to end-user criteria. However, few of these works were interested on how DEVS simulation code may be reused, structured, maintained, etc. In the following, we highlight these works related to our subject by showing their advantages and disadvantages. 

A first attempt to introduce design patterns in DEVS modeling and simulation software, was conducted by  \cite{Ferayorni2007}. They noted that the modeling capabilities of system theory do not support some important software design techniques and so without these kinds of software abstractions it is difficult to specify simulation models of complex software that include design patterns and explicit supports for the design of specific simulation modeling. The inclusion of design patterns in modeling and simulation environments such as DEVSJava for specific domains helps modelers and designers to develop software design specifications with key benefits: reduce cost and time of coding, enhance testing, etc. In \cite{Sarjoughian2015}, Sarjoughian et al. proposed a DEVS metamodel in order to structure and organize the modeling activity of developers based on Model Driven Architecture (MDA). The proposed Unified Modeling Language (UML) artifacts make easy the implementation of concrete models. However, we did not find how concrete models are mapped into runnable models.

James II is a DEVS software based on plug-in scheme, called the concept {\it plug'n simulate} \cite{Himmelspach2007}\cite{Himmelspach2008}. The  {\it plug'n simulate} architecture provides extensibility at two levels : (1) plug-ins types that define types of plug-ins by proposing an interface as an extension point and allowing concrete implementation and (2) plug-ins that provide different implementations of a given service (flatten versus hierarchical simulation, random number generators, etc.). This architecture allows any framework to integrate available and future plug-ins in a flexible and automatic process without modification of the existing code, based on the factory pattern. However, this solution imposes that each plug-in type should have an XML (eXtensible Markup Language) file defining some attributes useful for the factory. 

Dalle and Wainer \cite{Dalle2007} proposed modeling patterns for sharing components in a DEVS simulation. They developed the following patterns: the proxy, shortcut and matriochka patterns. The proxy pattern allows the modeler to share a component without creating new copies. So this pattern makes plugging of components in DEVS models possible. The second pattern shortcut allows building interactions between components. These interactions may be used to define bridges between components and may reduce the coupling complexity that DEVS often faces. Note that this pattern is useful in case of a layered architecture to create additional paths between layers and to comfort the traditional hierarchical interaction. The last pattern matriochka guarantees a safe interaction between the hierarchical DEVS components and enforces encapsulation.

Santucci and Cappochi \cite{Santucci2012} used design patterns to develop the extensible framework DEVSimPy. This framework claims extensibility by using the design pattern strategy to choose the best simulation algorithm at runtime, and using specific plug-ins based on the model view controller paradigm to start activity tracking during simulation. DEVSimPy provides other functionalities like extending a set of models with new behaviors as the blink plug-in which allows blinking a model making a transition during simulation, and updating the code of a  model during the simulation runtime. This last operation needs that the modeler stops the simulation, changes the code of a given model and restarts the simulation.

At first glance, the reader may believe that these works address the same goal of this work, ie., reusability and maintainability of DEVS simulation. However, by analysing these works, we are sure that they are not addressing the same problems and neither proposing the same solutions that we highlight in this paper. In fact, we attempt to provide patterns to how design and code DEVS models and do not tackle modeling problems that a modeler may meet in \cite{Dalle2007}. In addition, the technical solutions that support with the modeling patterns do not show how DEVS models may be designed except the fact the DEVS simulation may be designed through the Fractal architecture which can't answer our design requirements.

The {\it Plug'n simulate} concept that has been widely used by the simulation community allows uploading plug-ins in dynamic way. Consequently, simulation software such as James II are more flexible and extensible for further requirements. However, we did not find how this concept may be used to structure, design and code DEVS models. Unfortunately, James II does not provide inherent support for any modeling nor for any modeling paradigm  because all formalisms supported by this tool must inherit from the base classes of James II \cite{Himmelspach2007}.

In the rest of the paper, we propose to introduce design patterns for designing DEVS modeling requirements in order to reuse and maintain code of models. Note that due to the informal description of design pattern, we introduce OCL (Object Constraint Language) to the proposed patterns in order to make clear their interpretation.

\section{Definitions and recalls}
\label{stateDEVS}
\subsection{DEVS Formalism}
\label{DEVSform}
According to the literature on DEVS  \cite{Zeigler2018}, the specification of a discrete event model is a
structure given by:\\
$ M = (X, S, Y, \delta_{int}, \delta_{ext}, \lambda, D) $, where 
\begin{itemize}
	\item $X$ is the set of the external input events, 
	\item  $S$ the set of states, 
	\item $Y$ the set of the output events,
	\item $ \delta_{int}: S\rightarrow S$ is the internal transition function that defines
	the state changes caused by internal events, 
	\item $\delta_{ext}: Q\times X \rightarrow S$ is the external transition function that specifies the state
	changes due to external events,
	\item  $\lambda:S \rightarrow Y$ is the output function, and 
	\item $D: S\rightarrow R^+\cup\infty$ represents the maximum length or the lifetime function of a state. Thus, for a given state $s$, $D(s)$ represents the time during which the model will remain in state $s$ if no external event has occurred.
\end{itemize}

Zeigler~\cite{Zeigler1976} introduced the concept of the total states of a system: $Q = \{(s,e), s \in S \land e \in R^{+}, 0 \leq e \leq D (s)\}$, where $e$ represents the elapsed time in state $s$. The concept of total state is fundamental in that it permits one to specify a future state as a function of the time elapsed $e$ in the present state. There are potential benefits may lie in its ability to implement event filtering, wherein a planned change of state will be realized by a model only when the time interval that separates two key events exceeds a predefined value, and to encapsulate an otherwise mechanical event filtering at the conceptual level.

Thanks to the closure under coupling property of DEVS, atomic models are reusable using a DEVS coupled formalism that includes the specification of the
DEVS components and their couplings. The obtained model is defined by the following structure:\\
$MC = (X_{MC}, Y_{MC}, D_{MC}, M_{d \|d\in D},  EIC, EOC, IC, Select)$ where

\begin{itemize}
	\item$X_{MC}$: set of external events.
	\item$Y_{MC}$: set of output events.
	\item$D_{MC}$: set of component names.
	\item$M_d$: DEVS model named d.
	\item $EIC$: External Input Coupling relations.
	\item $EOC$: External Output Coupling relations.
	\item $IC$: Internal Coupling relations.
	\item $Select$: defines a priority between simultaneous events intended for different components.
\end{itemize}

\subsection{Object Oriented Design of DEVS Models}
In order to get a clear idea of existing design of DEVS models in OOP, we examined in research articles, technical documentation for DEVS software tools and packages,  and the given code examples, to highlight their designs in the form of class diagrams.

\begin{figure}[h]
	{
		\centering
		{\includegraphics[trim= 185mm 10mm 90mm 48mm, clip, height =9cm, width= 12cm]{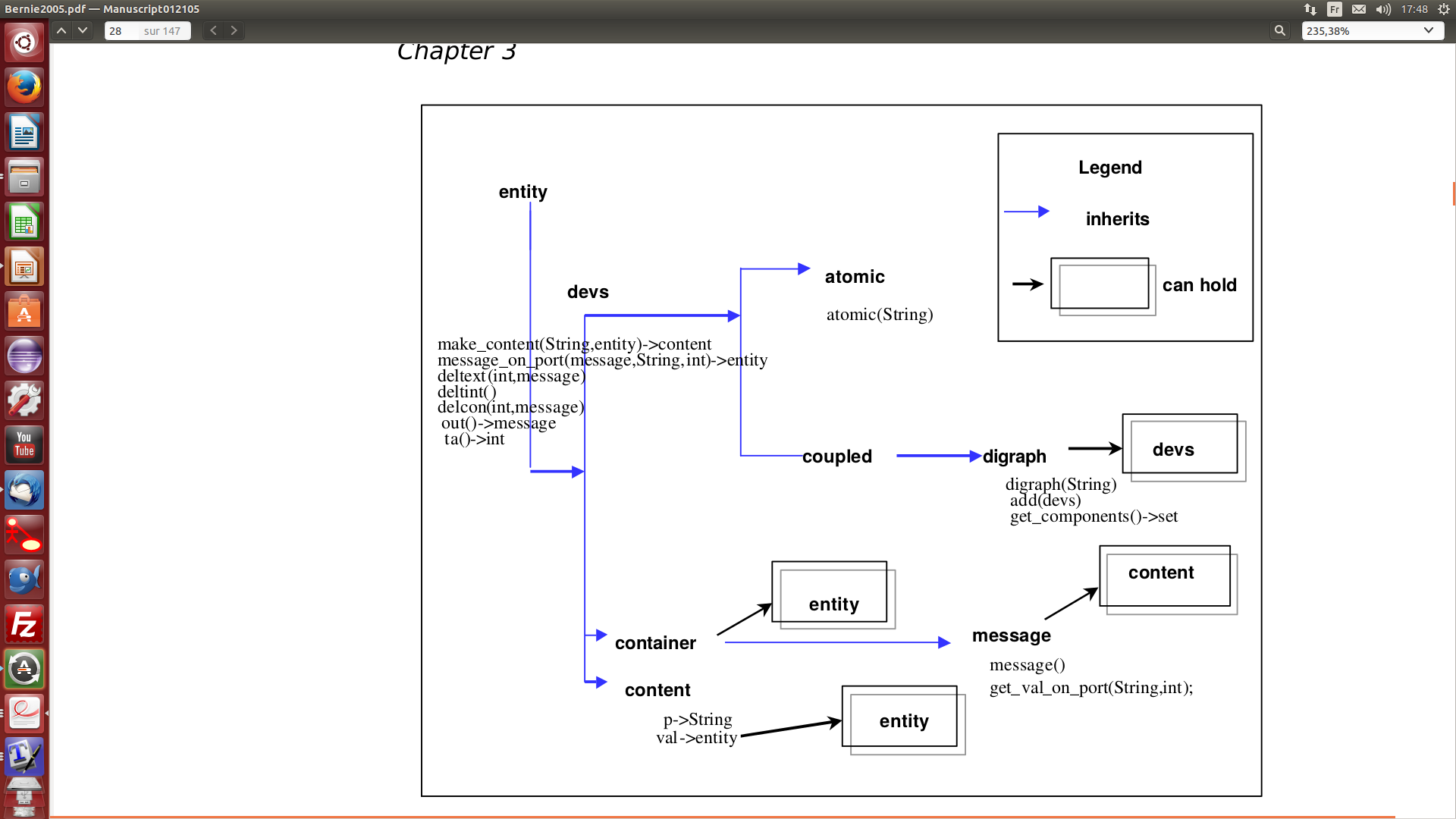}}
		\caption{The DEVSJava class hierarchy and main methods \cite{DEVSJAVA2012}.}
		\label{fig:devsjavafigure}
	}
\end{figure}

One of the well-known software in DEVS M\&S is DEVSJava. Its simulation kernel is based on Parallel-DEVS, so it is expected to design Parallel-DEVS models but can also simulate classic DEVS  by customizing the method {\tt deltext()}. The design of both Parallel- and classic DEVS models is done in a recursive way. Firstly, the designer extends an abstract class called {\tt atomic} to define for each DEVS atomic model its own class. Then, he extends the class {\tt digraph} to  create the instances of these new models which will be stored in a component set;  the designer defines the coupling between these instances using the method {\tt addCoupling(devs d1, String p1, devs d2, String p2)}. The method {\tt deltext(double e, message x)} provides to the model instance all external events  that occur from the simulator. However, while the designer motivation is  DEVS models, he should extend the class {\tt classic} to design his atomic models and should override the method {\tt deltext()} to handle one event at once; or a specific class {\tt siso} if the DEVS atomic model has only one input and one output ports and events are processed as reals. The class diagram of this description is shown in Figure~\ref{fig:devsjavafigure}.

\begin{figure}[h]
	{
		\centering
		{\includegraphics[trim= 65mm 45mm 50mm 80mm, clip, height =7.5cm, width= 0.99\linewidth]{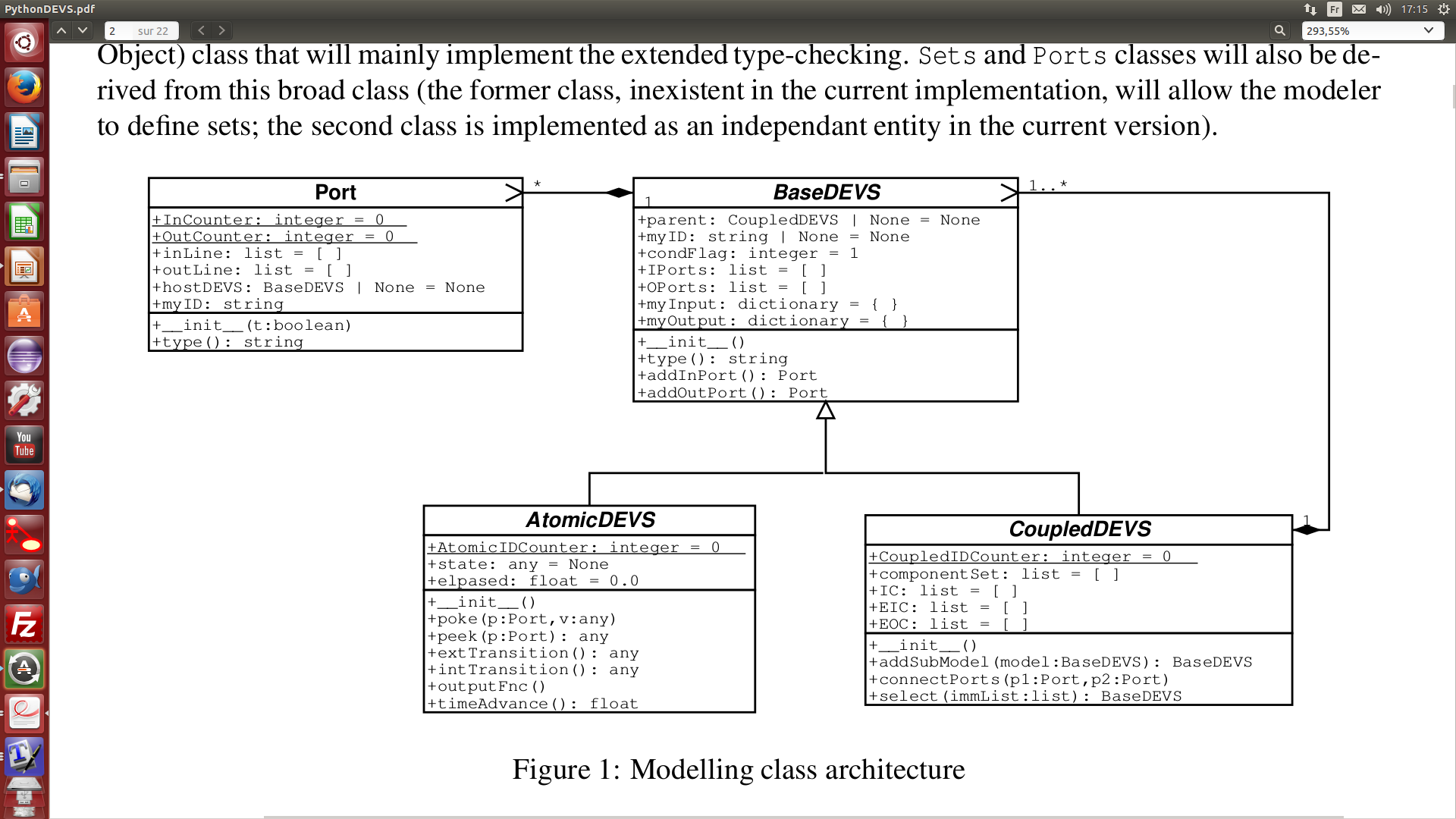}}
		\caption{The modeling archicture of PythonDEVS.}
		\label{fig:pythondesign}
	}
\end{figure}

PythonDEVS \cite{python2002} provides an interesting DEVS modeling package in which the designer should extend the classes {\tt AtomicDEVS} and  {\tt CoupledDEVS} in order to design DEVS atomic and coupled models respectively. These classes specialize a common abstract class {\tt BaseDEVS} which holds the lists of input and output ports. The class {\tt Port} may have the responsibility to conduct type checking when a coupling is made by verifying if the event sent out from the port sender is accepted by the port receiver. The class {\tt AtomicDEVS} acts as an interface that the designer extends to implement the different functions $\delta_{int}$, $\delta_{ext}$, $\lambda$, and {$D$} through the methods {\tt intTransition(), extTransition(), outputFnc() and timeAdvance()} respectively. In this class, there is a specific attribute {\tt state} used as a state variable. The designer may add other state variables in the inherited class to design the new atomic model. The class {\tt CoupledDEVS} allows designing new DEVS coupled models by inheritance. These new classes have the responsibility to store the instances of classes for their submodels {$D$} using the method {\tt addSubmodel(model: BaseDEVS)} and making the different coupling {$EIC, EOC, IC$} using the method {\tt connectPorts(p1: Port, p2: Port)}. The priority function {$select$}  that identifies the imminent model to handle when simultaneous events occur for a given coupled model, is specified by implementing the method {\tt select(immlist:list): BaseDEVS}. This package is shown on Figure~\ref{fig:pythondesign}.

The analysis of the DEVSJava and PythonDEVS class hierarchy diagrams  (cf. Figures~\ref{fig:devsjavafigure} and~\ref{fig:pythondesign} respectively) shows that there is a strong likeness to well-known design pattern: the composite pattern \cite{Gamma1995}. The use of design patterns in software development was justified in the literature on object programming through numerous research works and articles. Consequently, we believe more and more since our papers \cite{Hamri2010}\cite{Hamri2013} that some design patterns can be useful for efficiently designing simulations and especially DEVS  which in turn may enrich the design pattern library thanks to the coding experiences of DEVS designers.

\section{Design of DEVS Models}
\subsection{Designing DEVS atomic}
\subsubsection{Separating Structure from Behavior}
\label{sec:speratingstructurefrombehavior}
\begin{figure}[h]
	{
		\centering
		{\includegraphics[trim= 0mm 283mm 160mm 0mm, clip, height =4 cm, width= 8cm]{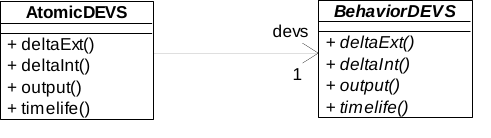}}
		\caption{Object design of atomic model.}
		\label{fig:atomicDesign}
	}
\end{figure}

As noted above, a very simple and efficient way to design an atomic model is to extend a class and implement the suitable behavior. In order to avoid such a design which is hard to maintain and update  due to the use of the inheritance relationship, we can use delegation to  replace inheritance in order to extract the behavior from structure then proceed in its implementation.  Fundamentally inheritance and delegation are dual~\cite{Stein1987}. Consequently we can favour inheritance or delegation according to our design requirements. 
Let us consider the design shown in Figure~\ref{fig:atomicDesign}.

From Figure~\ref{fig:atomicDesign}, the class {\tt AtomicDEVS} delegates each received message from the simulation to the attribute {\tt devs: BehaviorDEVS} which is responsible for updating the state variable and computing outputs. An example of delegation of the external transition function {\tt deltaExt()} is shown below:
\\
{\tt void deltaExt(Port p, Object ev, double etime)\{
	
	assert p.check(ev) // check optionally the recevability of ev by p
	
	devs.deltaExt(ev, etime);\\
	\}
}

About checking port type, Sarjoughian and Markid \cite{Sarjoughian2012} noted that the well-formedeness of port couplings between DEVS models is often ignored by DEVS simulation software; they proposed the use of EMF-Metamodeling to check  such a compatibility. However, we propose the use of object delegation to the class {\tt Port} to check whether an event may pass on a given port (inport or outport). Otherwise a runtime exception is thrown and the simulation is interrupted. The body of the method {\tt check()} depends on the type of events that a given port handles and for which the designer should code once the modeling requirements are known. Note that, the static and dynamic typing casts should be used carefully for checking receivability of events on a port. Indeed, the static and dynamic casts may convert a specific type into a common one and vice versa. In programming languages like Java and C++, many kinds of conversion are allowed for primitive types and classes. Considering primitives types, two kinds of conversion are possible widening and narrowing. Because, in DEVS, two interpretations of the couplings are given : 1) coupling in static way to connect DEVS models among them  to transmit events as such. In this case, the widening cast is advised; and 2) the coupling transforms events initially with a given type into another one. The designer should use or define the useful functions and methods in order to carry out the transformation and check whether the result is correct, in the case of coupling using functions.


On the other hand, when the set of events transiting a port is finite, we advise the use of enumerated sets allowing the definition of enumeration of events and making more safe the typing of events to their ports. 

The interface of the class {\tt BeahviorDEVS}  (it can be designed with an interface instead of a class) defines all methods that a given behavior should implement. It is a contract\footnote{The design by contract was introduced by Bertrand Meyer in his book  Object-Oriented Software Construction, in 1988. This approach defines a set of fundamentals for formal and precise specifications of software components.} that the designer should respect for any concrete DEVS behavior.

Note that using such a design, the designer may reuse the structure of DEVS atomic, ie.,  the class {\tt Atomic} to specialize it by adding new constraints and functionalities to the structure (add or delete a port); and may reuse the behavior through the classes implementing the interface {\tt DEVS} in other structures, i.e., other classes subtyping the class {\tt Atomic}.

\subsubsection{Designing DEVS Behavior}
Different designs may be conducted by designer according to the modeling and design requirements. However, we distinguish two main designs:
\begin{itemize}
	\item Classical design: A simple way to code a DEVS behavior consists of using the primitive types or enumerated sets to type events and states; and the conditional statement if-else or switch case to code state changes and compute output events. The logic of the conditional statement is well suited to the behavioral logic of DEVS, as shown through many examples found in the literature. Conditional statements using primitives types or enumerated sets also allows conducting fast simulations.
	\item Object design: If the use of OOP to design DEVS coupled is simple, in order to insure modularity and encapsulation of the code; it  is not  the case for designing behavior. A designer who wants to enhance reusability of DEVS behavior code, should objectify states, events and/or transitions. The object design is an alternative to the classical design in which the DEVS behavior is described via a compact block based on the conditional statement. In addition, such a design allows extending the objectified elements with new properties and operations using inheritance. Consequently, the passage from a DEVS design to an extended one with a new operational semantics (Parallel-DEVS, GDEVS, etc.), is guaranteed and simplified thanks to the object abstraction made on some DEVS elements.
\end{itemize}

\subsection{Designing DEVS coupled}

\begin{figure}[h]
	{
		\centering
		{\includegraphics[trim= 0mm 272mm 130mm 0mm, clip, height =5cm, width= \linewidth]{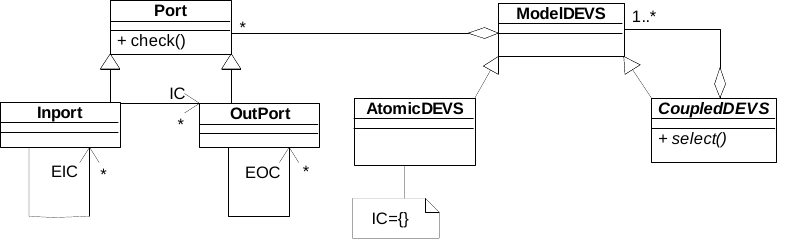}}
		\caption{Object design of coupled model.}
		\label{fig:coupledDesign}
	}
\end{figure}

A DEVS model may be coupled or atomic; and a coupled model is composed of at least one model. In addition, a model has a finite list of ports decomposed into two sets inports and outports. Bolduc and Vangheluwe~\cite{python2002} proposed an interesting design of DEVS which can be easily reused to design an extension of DEVS (see Figure~\ref{fig:pythondesign}). In fact, for an extension of DEVS which defines atomic and coupled models, the designer may makes its classes with the same structure. Consequently, we propose a design of coupled models shown in Figure~\ref{fig:coupledDesign} inspired by \cite{python2002} with slight modifications.

Note that this design is closer to the design pattern composite \cite{Gamma1995} which proposes to design hierarchical structure for a given object. This shows that some experiences in design patterns  in general may be reused by DEVS designers. The next section highlights techniques coming from software engineering and gives original solutions to design DEVS behaviors in OOP.

A designer may note that the design shown on Figure~\ref{fig:coupledDesign} suffers from inconsistency regarding DEVS semantics. For example, based on the aggregation between the classes {\tt CoupledDEVS} and {\tt ModelDEVS}, an instance of {\tt CoupledDEVS} may contain itself, a port may reference to itself, an output port belonging to an instance of {\tt AtomicDEVS} may reference an input port belonging to the same instance, etc. In order to remedy this within the semantics of UML class diagrams, we propose the use of OCL (Object Constraint Language) to complete this class diagram, even if notes are placed on summarizing constraints that the designer should respect. 

An interesting feature of this class diagram is that extensions of DEVS are feasible from classes {\tt ModelDEVS}, {\tt CoupledDEVS} and {\tt AtomicDEVS}. Consequently, these classes may have a lot of subtypings (instances in DEVS and its extension). In order to guarantee the homogeneity of submodels to the same formalism (DEVS or a given extension and not both), a verification on type of the submodel should be conducted before adding it to coupled model. Unfortunately, OOP does not provides mechanisms to guarantee  such a constraint but it could be added to the design.

\section{Object Design of DEVS Behaviors}

Designing DEVS behaviors using object oriented paradigm and getting an object-oriented code requires a particular attention on designing discrete event models in general. People proposed a lot of designs for statecharts, Petri-nets, DEVS, etc. Most of them propose to objectify states and transitions. However, the dynamics between states and transitions uses  conditional statements or structures like tables. On the other hand, events are often typed with primitive types. Such designs lead to some disadvantages (that we try to remedy in our framework) listed below:
\begin{itemize}
	\item the dynamics expressed using conditional statements is hard to maintain and is impossible to update at run time.
	\item the no objectifying of events leads to code less abstract. In fact designing  events with objects (when it is necessary) allows redefining the event concept easily (for example, assign to the time occurrence a temporal window instead of a unique value).
\end{itemize}
The next subsections discuss some solutions that may be helpful for the designers of DEVS behaviors.

\subsection{State Design Pattern}
The state design pattern \cite {Gamma1995} and its variants \cite{Dyson1998}\cite{Adamczyk2004}\cite{Chine2010} \cite{Garcia2013}  allows designing dynamic objects in which the behavior depends on the state. The state pattern consist of an abstract class {\tt State} holding all events acting on the object (finite state machine) as public abstract methods. For each state of the object a subclass is created and event methods are implemented which act as transitions and return the future state (or update it directly). The object to design holds a reference on the current state and all received events are delegated to this reference in order to make a transition and update its current state.

\begin{figure}[h]
	{
		\centering
		{\includegraphics[trim= 0mm 268mm 130mm 0mm, clip, height =5.5cm, width= \linewidth]{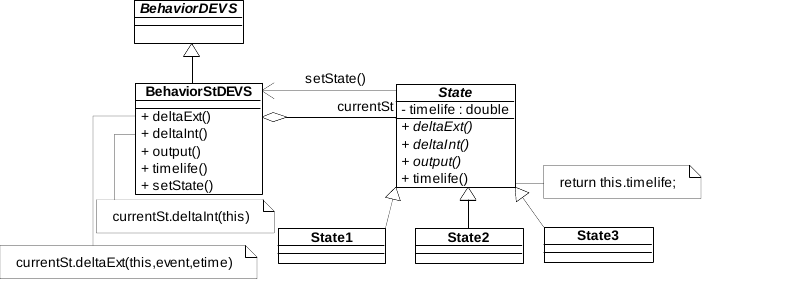}}
		\caption{Design of DEVS behavior using the state pattern.}
		\label{fig:statedesign}
	}
\end{figure}

Using the state pattern, to design a DEVS behavior (see Figure~\ref{fig:statedesign}), requires identifying for each state the possible handled events, and expressing the computation of the next state using conditional statement over the possible events. The second solution consists of identifying by the class {\tt BehaviorDEVS} which method to invoke of the current state, by using reflection. Note that, the state object is responsible for updating the current state. From this fact, the state pattern is a partial solution to make code more abstract and reusable because the designed behavior remains compact without objectifying events, even if it is divided and dispatched on different state objects. The following subsections discuss  more elaborated solutions.

\subsection{State Event Design Pattern}
Some design patterns of behavioral objects attempt to objectify events in addition to states \cite{Fowler1999} \cite{Adamczyk2004}. However, the given solutions are not completely satisfactory. \cite{Fowler1999} solution consists of designing event communication between objects. On the other hand, \cite{Adamczyk2004} solution, even if states and events are objectified, the dynamics through state changes is expressed using conditional statements. This causes a loss of maintainability through the use of conditional statement and a loss of performances due to objectifying events and states.  

In the state event pattern \cite{Hamri2014} \footnote{For a complete documentation of the state event pattern, the reader may accede to the author article available on ACM digital library (DOI: 10.1145/2721956.2721987)}, we proposed to objectify events in addition to states and we used delegation to event objects in order to design state changes. The state event pattern designs DEVS behaviors expressed with a finite set of events. State objects may be declared from the class {\tt State} or a subclass for each one extending this class. On the other hand, events are designed by extending the class {\tt Event}. Note that for each received event, the atomic model which holds a reference on its current state, delegates the state change to the occurred event; and each event object holds all its state changes through the attribute {\tt transitions: HashMap(State, State)}. A specific event subclass {\tt INTEVENT} is designed a priori in order to hold internal transitions caused by autonomous changes of the system. The body of the method {\tt deltaExt()} belonging to the class {\tt DEVSBehavior} which is not responsible for updating the current state is as follows:
\\
{\tt void deltaExt(Object event, double etime)\{
	\\
	((Event)event).setChange(this);
	\\
	\}
}

However, this way of designing transitions supposes that only one fireable transition exists for a given current state and for the occurred event. In case of stochastic models, an event may have different transitions for a given state, so the attribute should be rewritten as {\tt transitions: HashMap(State, List<State>) } to hold a list of possible future states for a given event.

\begin{figure}[h]
	{
		\centering
		{\includegraphics[trim= 0mm 260mm 115mm 0mm, clip, height =6.5cm, width= \linewidth]{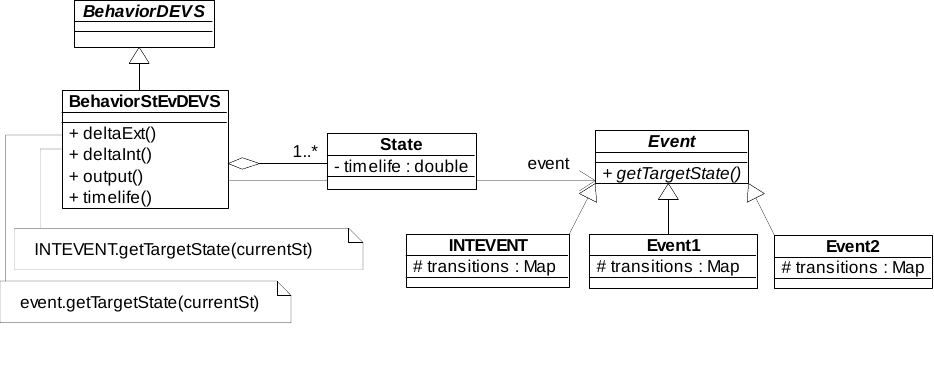}}
		\caption{Extension of the state event pattern to DEVS behavior.}
		\label{fig:stateevent}
	}
\end{figure}

The discussed solution is summarized through the following class diagram shown on Figure~\ref{fig:stateevent}, after having adapted the state event pattern to DEVS.

Moreover state objects should be declared as unique to get a unique reference for each state. This constraint may be insured by the class {\tt DEVSBehavior} or by using the Singleton pattern when state objects are subclasses of the class {\tt State}. Note that in UML, a static attribute is shared between all objects assigned directly from a given class or its subclass. In order to avoid  event objects sharing the same transitions, the attribute {\tt transition} is declared inside each subclass {\tt Event} as static and not a common attribute in the class {\tt Event}. A serious constraint of the OOP to make design more clear and synthetic.

\subsection{State Event Transition Design Pattern}
\begin{figure}[h]
	{
		\centering
		{\includegraphics[trim= 0mm 130mm 5mm 0mm, clip, height =5.5cm, width= \linewidth]{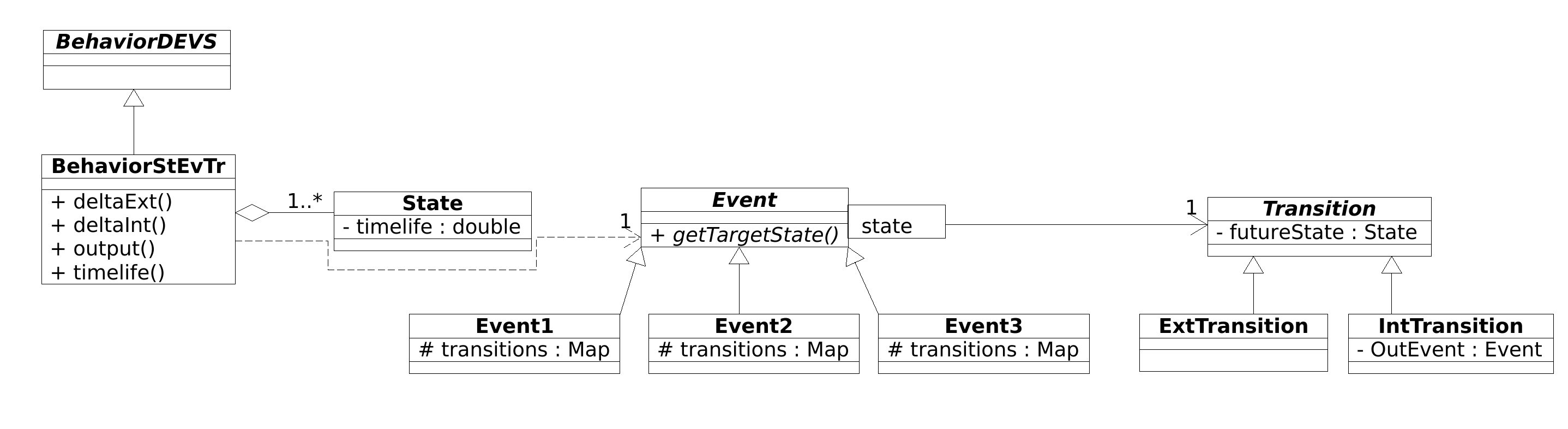}}
		\caption{State event transition pattern extended to DEVS behavior.}
		\label{fig:stateeventtransition}
	}
\end{figure}

This pattern (Figure~\ref{fig:stateeventtransition}) is an extension of the state event one, in which we objectify the element transition of DEVS using the class {\tt Transition}. We use delegation to identify the fireable transition from the occurred event object and the current state of the model. So, this transition returns the next state of the model ({\tt futureState}) allowing to set a new state. From the abstract class {\tt Transition} are defined two classes {\tt ExtTransition} and {\tt IntTransition} to design external and internal transitions respectively. Instances from both classes are stored in each event class to hold its own possible transitions.  Note that the use of delegation leads to a fair coupling between used objects (state, event and transition objects).

One of benefits of this pattern is that the class {\tt Transition} may be overridden with new data and methods. In fact, it is easy to design a transition with a condition that should be checked to make a given state change and an action to update some state variables; always thanks to the OOP which allows preserving the other classes ({\tt State}, {\tt Event}, etc.).

\section{Formalizing DEVS designs to Get Consistent Models}
DEVS class diagrams shown on Figures \ref{fig:atomicDesign}--\ref{fig:stateeventtransition} are incomplete because the UML language is not formal and ambiguous. Zinoviev~\cite{Zinoviev2005} proposed a set of formalized rules to map DEVS into UML. However these rules are specific to Parallel-DEVS and should be rewritten to be integrated directly in an executable framework. On the other hand, the given object-oriented designs of DEVS enrich that discussed in \cite{Zinoviev2005}  leading us to describe other rules.

OCL\footnote{OCL does not support describe constraints on multiple inheritance and infinite sets of object like {\tt Intger.allInstance()}} (Object Constraint Language) \cite{Warmer2003} is a standard language of UML to describe constraints on objects showing how they should behave and for which a given class diagram can not explicit such constraints. In other terms, OCL allows describing a semantics for class diagrams. In addition, OCL is supported by many software tools  allowing an automatic translation of OCL constraints to compilable code (Java, C++, etc.) and so adopting clearly and formally the well-known concept of design by contract.

Now let us consider the following OCL constraints for DEVS:
\begin{enumerate}
	\item Avoiding feedback loop couplings in DEVS atomic.\\
	{\tt context ModelDEVS \\
		inv : self.outports$\rightarrow$forAll( p : Port, p.influencee\\ $\rightarrow$intersection(self.inports)$\rightarrow$isEmpty())
	}\\
	
	this constraint allows a given atomic model influencing only another port belonging to a different model.
	\item Distinguishable inport and outport sets.\\
	{
		\tt context ModelDEVS\\
		inv : self.inports$\rightarrow$intersection(self.outports)$\rightarrow$isEmpty()
	}\\
	this constraint guarantees that a port for a given DEVS model can not be both an inport and an outport.
	
	\item Defining a correct set of submodels for a DEVS coupled.\\
	{
		\tt context CoupledDEVS\\
		-- a coupled model reuses at least one submodel\\
		inv : self.submodels$\rightarrow$notEmpty() \\
		-- a coupled model does not be a part of its submodels\\
		inv : self.submodels$\rightarrow$excludes(self)\\ 
	}\\
	these two constraints defines that a given DEVS coupled model contains at least one DEVS model (atomic or coupled) and the model itself will not be part of its submodels respectively.
	\item Checking compatibility between states, transitions and lifetimes.\\
	According to the design chosen for DEVS, a set of constraints should be defined. The object design that objectifies only states leads to the following constraints:\\
	{\tt context State\\
		inv: self.isTypeOf(ActiveState) $\equiv$ self.timeLife() < $\infty$\\
		inv: self.isTypeOf(activeState)  $\equiv$ not self.isTypeOf(passiveState) 
	}\\
	
	However, for the use of a classical design of DEVS, consider the following constraint:
	{\\
		\tt context BehaviorClassicDEVS\\
		inv: self.isPassive() $\equiv$ self.timelife() $= \infty$ \\
		inv: self.isPassive()  $\equiv$ not self.isActive()
	} \\
	
	OCL allows also defining other constraints on objects in addition to invariants like init attributes, pre- and post-conditions of methods, etc. So, the designer may apply these constraints to DEVS while updating state variables of the model, as:
	\item Guaranteeing the lifetime of any state is positive with the following constraint post-condition.\\
	\hspace{2.5 cm}
	{\tt
		context BehaviorDEVS::timelife()\\
		\hspace{2.5 cm}
		post :  result $\geq$ 0
	}
	
\end{enumerate}

With these DEVS OCL constraints, we guarantee that the design of a given model has a correct syntax and is consistent. The framework can check, in run mode, whether the constraints are respected or violated if they are incorporated into the final code.

Note that these constraints are specific to DEVS. In case of  DEVS extension, other constraints may be described according to the new extension requirements that should not violate the OCL constraints of DEVS. In fact the DEVS constraints should be respected in the new formalism, otherwise it is not considered as an extension. 

On the other hand, OCL  constraints may be used to define assertions for modeling requirements of a given system. In DEVS counter example ~\cite{Zeigler2000}, the following constraint insures that the counter value has always a positive or a null value:\\
{\tt context Counter\\
	inv:  self.Counter $\geq$ 0\\
}

\begin{figure}[t]
	{
		\centering
		{\includegraphics[trim= 5mm 35mm 10mm 10mm, clip, height =8cm, width= 12cm]{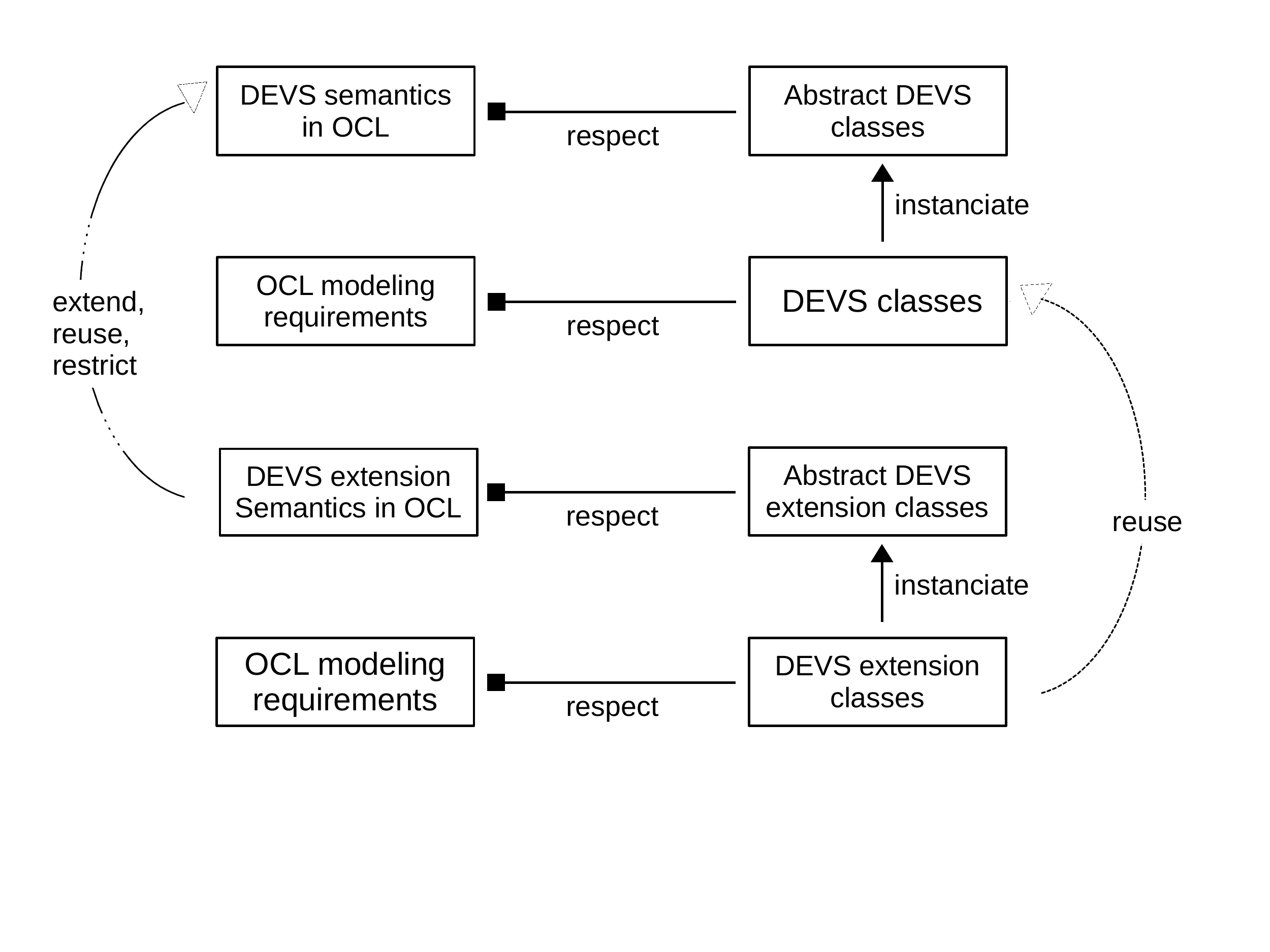}}
		\caption{OCL framework compliant to DEVS.}
		\label{fig:ocl-devs}
	}
\end{figure}

In general, OCL rules should be defined at formalism and model levels in order to guarantee a well formedness of models and implementations to DEVS (see Figure~\ref{fig:ocl-devs}).

The different OCL rules written above are given to the designer who may reuse, extend or restrict for customizing them and check the consistency of  extensions or designed models. Let us consider Real Time DEVS (RT-DEVS) \cite{Hong1997} an extension to DEVS in which the lifetime of each state is an interval.  Checking whether the lifetime of any state in RT-DEVS has a positive time interval can be inferred from the OCL rule (5) of DEVS as follows:

$\forall s \in S_{rt-devs} \exists \alpha, \beta \in R$

$\rightarrow   ta_{rt-devs}(s) = [\alpha, \beta] $  $ ta_{rt-devs}\xrightarrow{\text{extend}} ta_{devs}$

$\rightarrow   ta_{rt-devs}(s) = [\alpha, \beta] $  $ ta_{rt-devs}\xrightarrow{\text{extend}} ta_{devs}$ post $ta_{devs}() \geq 0$

$\rightarrow   ta_{rt-devs}(s) = [\alpha, \beta] $  post $ta_{rt-devs}() \geq 0$

$\rightarrow 0\leq \alpha \leq  ta_{rt-devs}(s)\leq \beta  \rightarrow \alpha \in R^{+}$   $ \beta \in R^{+}$

which corresponds to the lifetime function definition in RT-DEVS.

Moreover, this shows that any state in RT-DEVS  has a positive time interval based on the definition of DEVS lifetime function. Note that other properties of the extension may be checked from DEVS using logics and inferences.

\section{Reusability and Maintainability of DEVS Models}

Few studies on the maintainability of DEVS models are published. Perhaps, the main reason is that DEVS community is not interested on maintaining simulation models and this is viewed as software engineering requirement and  challenge. Note that in DEVS, reusability is explored only through the simulation engine and libraries of models. So what about adding new behaviors to atomic models?  What about specializing a library of DEVS models to a new extension? The following subsections give answers to these two questions (even if we answer the second question succinctly).

Essentially, reusability of DEVS (extensions) models consists of reusing both atomic and coupled models to create new models. This supposes that once the atomic model produces an acceptable behavior, it will be encapsulated in a box with ports and reused such as to design new models. In some cases, a new model is made from an existing one by adding and/or deleting its submodels, ports, states, etc.  and both models are saved (two models will exist).  Even if at the conceptual level that is feasible but how it can be done at code? This requires extending the classes of a given model to create new classes according to the requirements of the new model and overriding methods yet implemented to express new behaviors of $\delta_{ext}$, $\delta_{int}$, etc. Note that we can reuse these methods  but we can\rq{}t update their bodies. Recall that in OOP reusability is favored by using inheritance and delegation, and encapsulating each part of the object subject to change in class.

In simulation, DEVS models are manipulated through object instances made from classes describing behaviors verified and validated by the designer. Consider a designer formulates new modeling requirements for existing models. Consequently, new classes should be designed either from scratch (this is a bad solution from software engineering view); or from existing DEVS classes (that is we are considering). In fact, reusing existing DEVS classes  to design new classes, i.e, new models is possible only if the preliminary (basic) DEVS  classes are maintainable, i.e, ready for easy modifications. Otherwise the basic DEVS classes are inflexible and less reusable.

\begin{figure}[h]
	\begin{center}
		\subfigure{	    
			\includegraphics[trim= 5mm 120mm 145mm 0mm, clip, width = 5.5cm, height=3cm, scale = 0.2]{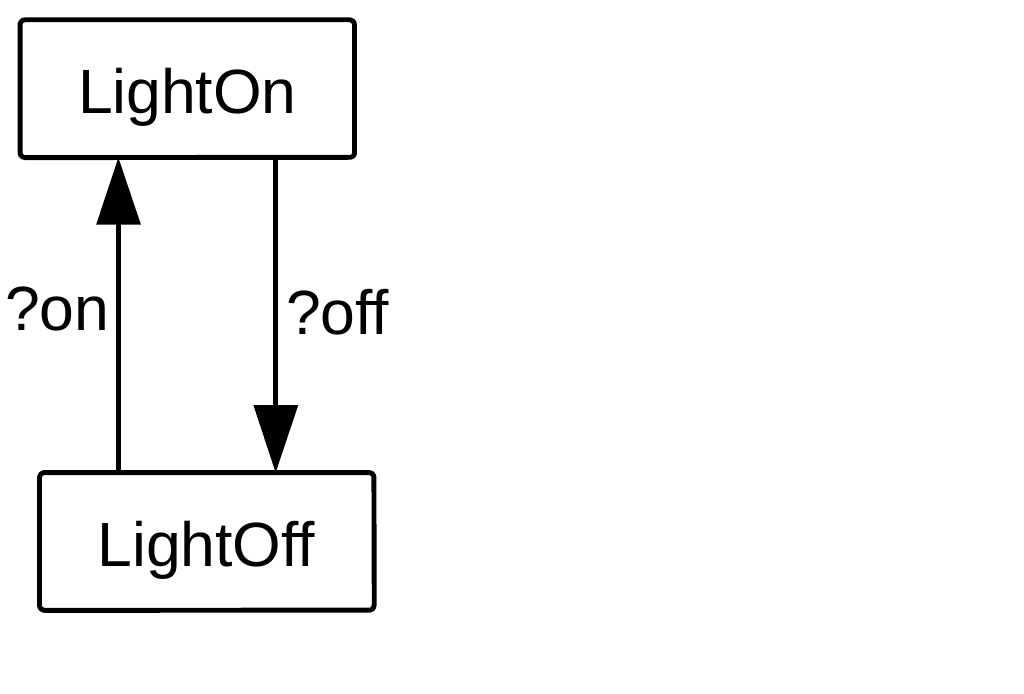}
		}		 	
		\subfigure{
			\includegraphics[trim= 2mm 120mm 125mm 0mm, clip, width = 5.5cm, height=3cm, scale = 0.2]{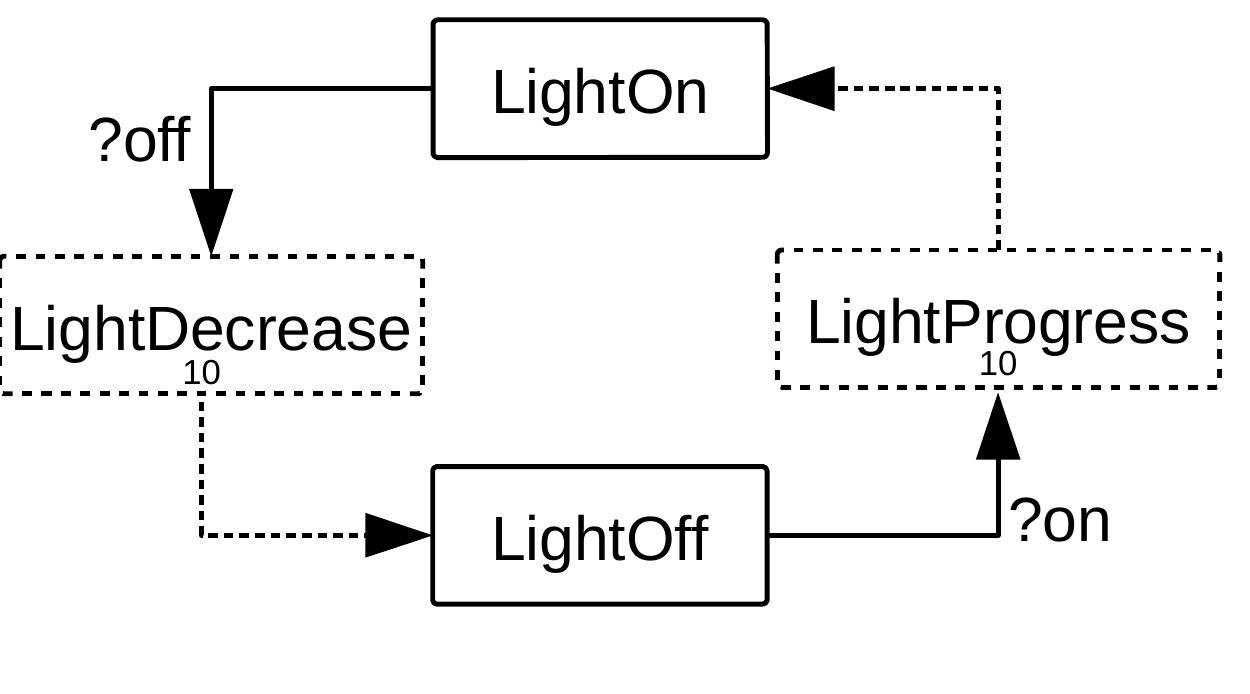}  
		}
		\caption{DEVS behaviors of lamp : static (on the left) vs variable (on the right) light.}
		\label{fig:lightlamp}
	\end{center}
\end{figure}

Now let us consider two different light lamps (see Figure~\ref{fig:lightlamp}). The first one has a static light, the second one provides a progressing and decreasing light before switching on and off respectively.

Comparing the two states models shown on Figure~\ref{fig:lightlamp}. The lamp with light variation reuses the classic lamp behavior. In fact, we add to the first model two actives states (LightProgress and LightDecrease) and two internal transitions (LightProgress $\mapsto$ LightOn and LightDecrease  $\mapsto$ LightOff); and we update its external transitions (LightOff $\mapsto$ LightProgress and LightOn $\mapsto$ LightDecrease). Is it possible to do that on the executable code of the classic lamp model? If the model is coded with one class in which the behavior is a bloc of switch case statements, the answer is no. 

\begin{figure}[h]
	{
		\centering
		{\includegraphics[trim= 5mm 225mm 5mm 5mm, clip, height =5cm, width= \linewidth]{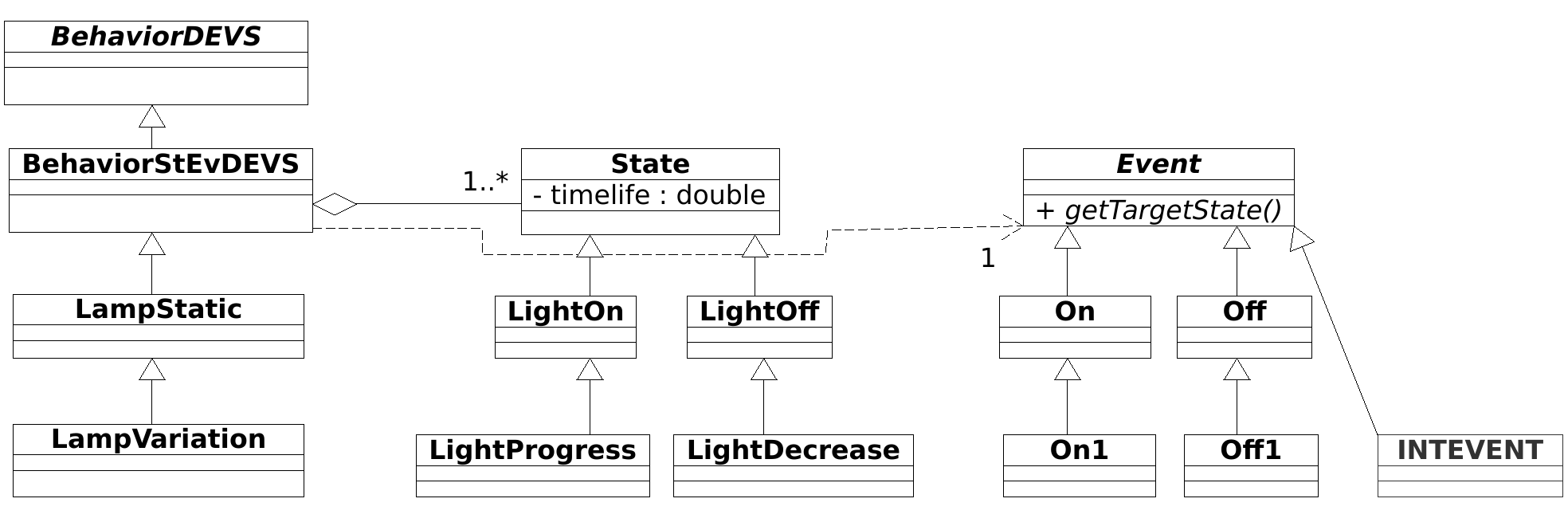}}
		\caption{Class diagram of the lamps: static and variable lights.}
		\label{fig:lamp-class diagram}
	}
\end{figure}

However, by objectifying the states and events of the classic lamp model, the modeler can easily design the lamp with light variation by adding two state classes {\tt LightProgress} and {\tt LightDecrease} and two event classes {\tt On1} and {\tt Off1} to redesign the event on and off by inheritance from the old classes {\tt On} and {\tt Off} . Figure~\ref{fig:lamp-class diagram} shows the classes invoked in the design of two lamps with a static and variable light.

Note that, the structure of the classic lamp model is reused to design the structure of the new lamp, i.e., the class {\tt LampVariation} inherits the class {\tt LampStatic}. In addition two state classes are reused {\tt LightOn} and {\tt LightOff}, and two event classes are extended to design the event classes {\tt On1} and {\tt Off1} from the existing event classes {\tt On} and {\tt Off}. The class {\tt INTEVENT} is added to design autonomous changes

An interesting feature of such a design is that the review of the executable code is easily conducted. Thanks to the reification of states and events the designer may easily deduce how each object should behaves according to the described model. 

However, such a design suffers from a lot of classes to design and manage, unlike a classical design in which there is only one class in order to design a DEVS behavior but this the cost of enhancing reusability and maintainability of code. In any case, separating structure of DEVS from its behavior and objectifying states, events and transitions are based on the OOP principle which consists of encapsulating data subject to modification in independent classes.

\section{Comparison of Performance of Different DEVS Designs}

\subsection{Using System Performances and Resources}
In this experimental study, we are interested in two parameters: time of execution and size of memory heap for each design applied to a model $M$ made automatically that consists of $n$ states, $m$ events and $n\times m$ transitions. Every one knows a priori that an event-driven design using a conditional statement is faster than the object-oriented one. In fact, in the classic and popular design there is no push-pop memory to carry out a state change. However, the state, state event and state event transition patterns carry out one, two and three push-pop memory respectively, to make one state change (cf. the transition methods of each design). Obviously, this needs some more memory bytes and slows down the simulation. For a seek of simplicity, we limit this study to a small set of states ($n = 2..5$) and events ($m=2..10$).

This study is conducted on personal computer DELL with CPU Intel  Core2 Duo CPU E8400 - 3.00GHz $\times$ 2 where is set Ubuntu 14.04. In order to carry out such simulations on different code according to the proposed DEVS designs, we programmed a module to instantiate randomly the behavior of the model $M$  and we developed a system to generate the source code of the given model $M$ according to each DEVS design pattern. In all given object designs, reflection is used to map random events  to objects. Finally  simulations are run and the runtime mean of each design is given on Table~\ref{table:perf}.

\begin{center}
	\begin{table}[h]%
		\centering
		\caption{Mean runtime in millisecond of the model $M$.\label{table:perf}}%
		\begin{tabular}{p{1.8cm}p{1.7cm}p{1.8cm}p{2.3cm}p{2.5cm}} 
			\toprule
			DEVS design & Switch case & State pattern & State event pattern & State event  transition pattern\\
			\midrule
			Mean runtime & $0.025$ & $0.315$ & $0.953$ & $1.159$\\
			\toprule
		\end{tabular}
	\end{table}
\end{center}
The shown results in Table~\ref{table:perf} confirm that faster discrete event simulations are those designed with the switch case statement. The cost of objectifying state, event and transition increases with the number of elements to objectify.  The state pattern provides an execution time lower than the state event pattern, which in turn provides an execution time lower than the state event transition pattern. In any case, the execution time of simulation designed with object pattern remains less than few milliseconds.

In return, the given object design patterns  are able to design event-driven behavior with a large size of events and states than the classic design. In object-oriented programming like Java and C++, the file size for a given class should not exceed the maximum configured. Consequently, in this experiment, the classic design exceeds first the file size authorized for a Java class, then the state pattern and finally the state event and state event transition patterns. In fact, these patterns provide a systematic approach to dispatch code through different classes.

Recall that the different designs given in this work produce the same behavior. This reinforces that object designs of DEVS simulation are able to conduct correct simulations.

In conclusion, the need of quick simulation pushes the designers to use the conditional statement to design event-driven behaviors. However, the given object patterns enhance the reusability and maintainability of such behaviors. Unintentionally,  the proposed patterns design behaviors with a large size of states and events.

\subsection{Experimenting the DEVS patterns in Simulation}

The  core of this work consists of proposing object-oriented designs for DEVS modeler. Thus, it is necessary to measure the reusability and maintainability using metrics for each design.  Well-known object-oriented metrics is proposed by \cite{Chidamber1994} \cite{Cheikhi2014}, called C\&K metrics, and that were created in order to reinforce such  measurements and consolidate decisions from this metrics.

The C\&K metrics consists of six measurements: weighted methods per class (WMC), depth of inheritance tree (DIT), number of children (NOC), coupling between object classes (CBO), response for classes (RFC) and lack of cohesion in methods (LCOM) for each class. Moreover this metrics, we add two additional measurements: the number of classes (NOC$_l$) and the total line of code (TLOC) which allows us to compare efficiently the proposed DEVS designs.

\begin{figure}[!h]
	{
		\centering
		{\includegraphics[trim= 0mm 105mm 30mm 1mm, clip, height =6cm, width= \linewidth]{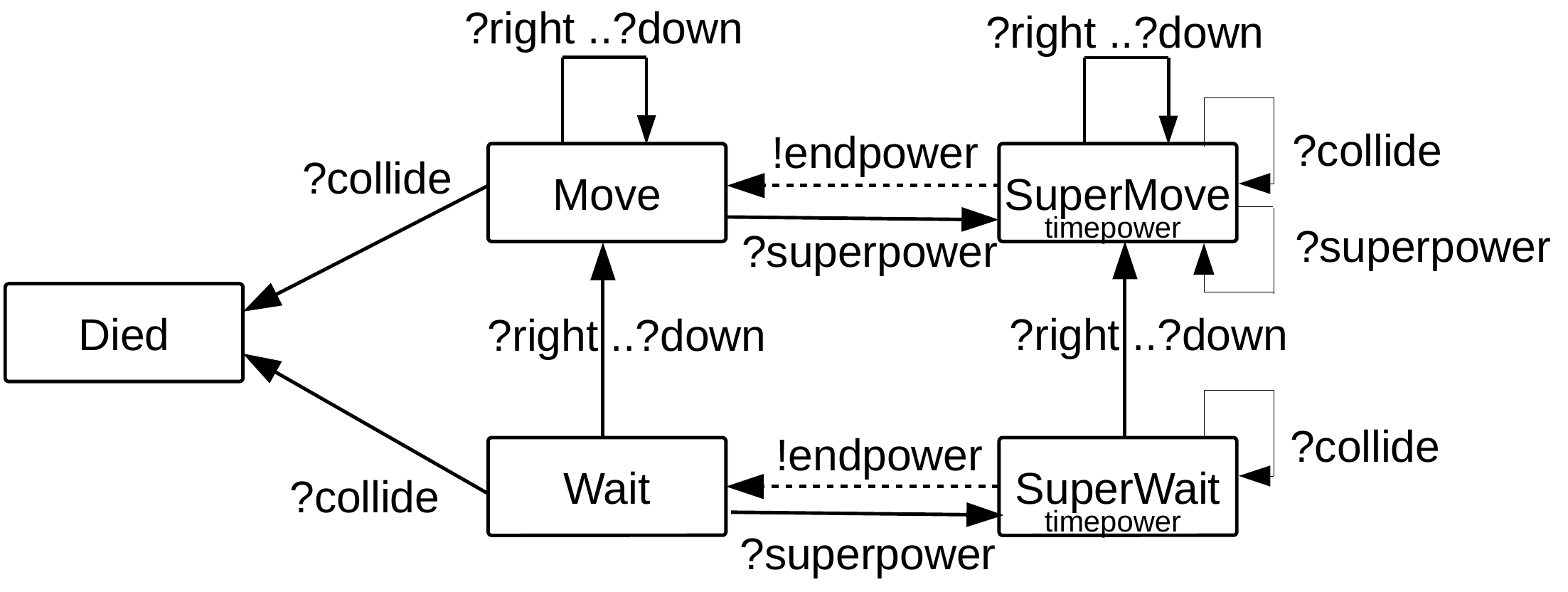}}
		\caption{DEVS behavior of Pacman.}
		\label{fig:pacman}
	}
\end{figure}

Now, let us consider the game Pacman which is used in different research works for two decade to study fields like robotics, biology, sociology, psychology and the most active field computational intelligence \cite{Rohlfshagen2018}. In DEVS, some works used Pacman as a case study to validate a given approach or some results \cite{Gore2015, Syriani2013}. From us, we want to check whether or not the described behavior for Pacman is correct.  We consider Pacman with 4 movements (left, right, up and down) useful to travel inside the labyrinth and to eat all pac-dots met; and  two kinds of collisions, one with the walls of labyrinth and a second with the enemies.  If Pacman eats a super pac-dot, then it will activate its super power to beat the enemies and will move more quickly for a given time. The corresponding DEVS model is shown on Figure~\ref{fig:pacman}.

Consider the behavior of Pacman designed in the four designs: switch-case (1), state pattern (2), event pattern (3) and state event transition pattern (4). The different simulations of the Pacman based on these designs were carried out successfully and produced the expected behavior, i.e., all the simulations fired the right state change at the right time. Consequently, we may assume that the designs of Pacman are valid, they produce the same behavior and a priori they are free from errors. 

In order to evaluate the reusability and maintainability of the different designs for the Pacman entity, we use the plug-in metrics 1.3.6 \copyright~ for eclipse \copyright~ projects. So, let us consider a significant subset of measurements selected from those provided by the plug-in, shown in Table~\ref{table:measurepacman}.

\begin{center}
	\begin{table}[h]%
		\centering
		\caption{Metrics of the Pacman DEVS designs.\label{table:measurepacman}}%
		\begin{tabular}{lccccc}

		\toprule
		
		\multicolumn{1}{c}{}               {\bf Metrics  }                 &                             & {\bf  (1) } & {\bf  (2)} & {\bf  (3) }&  {\bf  (4)}\\ \midrule
		\multicolumn{2}{c}{NOC}  & 5           & 11            & 20            & 34  \\ \midrule
		\multicolumn{1}{c}{{Abstractness}}              & avg.             & 0           & 0.087         & 0.093         & 0.087                                                                    \\ 
		\cline{2-6} 
		\multicolumn{1}{c}{}                                           & max              & 0           & 0.167         & 0.167         & 0.167                                                                    \\  \midrule
		\multicolumn{2}{c}{TLOC}                                                          & 234         & 415           & 353           & 515                                                                      \\  \midrule
		\multicolumn{1}{c}{DIT}                       & avg.             & 2.6         & 2.182         & 20.5          & 2                                                                        \\ \cline{2-6} 
		\multicolumn{1}{c}{}                                           & max              & 3           & 3             & 3             & 3                                                                        \\ \midrule
		\multicolumn{2}{c}{Num. of Packages} & 1           & 2             & 3             & 4                                                                        \\ \midrule
		\multicolumn{1}{c}{LCOM}                      & avg.             & 0.092       & 0.076         & 0.042         & 0.042                                                                    \\ \cline{2-6} 
		\multicolumn{1}{c}{}                                           & max              & 0.333       & 0.5           & 0.5           & 0.5                                                                      \\ \midrule
		\multicolumn{2}{c}{NOCh}                                                          & 0           & 5             & 13            & 26                                                                       \\ \midrule
		\multicolumn{1}{c}{WMC}                       & avg.             & 12.2        & 8.909         & 2.8           & 2.382                                                                    \\ \cline{2-6} 
		\multicolumn{1}{c}{}                                           & max              & 46          & 15            & 8             & 10                                                                       \\ \midrule
		\multicolumn{2}{c}{Instability}                                                   & 1           & 0.5           & 0.568         & 0.568                                                                    \\ \midrule
		\multicolumn{1}{c}{MCCC}                       & avg.             & 7.667        & 2.5         & 1.333           & 1.667                                                                    \\ 
		\cline{2-6} 
		\multicolumn{1}{c}{}                                           & max              & 34          & 10            & 3             & 3                                                                       \\ 
		
		\toprule

		\end{tabular}
		
	\end{table}
\end{center}

According to the interpretation of object-oriented metrics given in the literature, we make the following remarks and conclusions: the number of classes (NOC) increases from a less object-oriented design until a full object-oriented one thanks to the objectifying of states, events and transitions. This is a good thing for the readability criteria of the code.  The measurement Total Line Of Code (TLOC) increases significantly from the design (1) to (4), this is due to the additional methods  as the setters and getters of classes. The measurement Deep Inheritance Tree (DIT) is stable for each design, this is due to use of the composite design pattern and the separation of the behavior from the structure of DEVS atomic models. The measurement Lines Code Of Methods decreases from the design (1) to (4), this shows formally that the attributes and methods of classes in object-oriented design are more cohesive i.e., more unified.  The measurement Weighted Method Class (WMC) decreases substantially in the designs (3) and (4), this shows clearly that the design patterns event and state event transition decrease the complexity of classes.

On the other hand, the important measurement, the Mac Caby Cyclomatic Complexity (MCCC) which measures the readability, debugging and maintainability of code \cite{McCabe1989}, decreases significantly from the design (1) to (4). Based on these measurements, we conclude that object-oriented designs for DEVS are useful for designers to make easy the reusability, readability, test and maintainability of code of models.

\subsection{Desgining Functions with Patterns}

Functions are a simple way to express complex operations on finite and infinite spaces. In addition they are easy to design using conditional statements. However, it is not obvious to design them in object paradigm without making abstractions.

Many programming languages like Java provides representing numbers using objects. Even if using objects consumes more memory and slows down computation comparing to primitive types, it has some advantages like  allowing the use of the {\tt NULL} constant to note that a number is not initialized instead of a default value or declaring collections of numbers which is impossible using primitive types.

\begin{figure}[h]
	{
		\centering
		{\includegraphics[trim= 0mm 160mm 5mm 0mm, clip, height =1.5cm, width= 14cm]{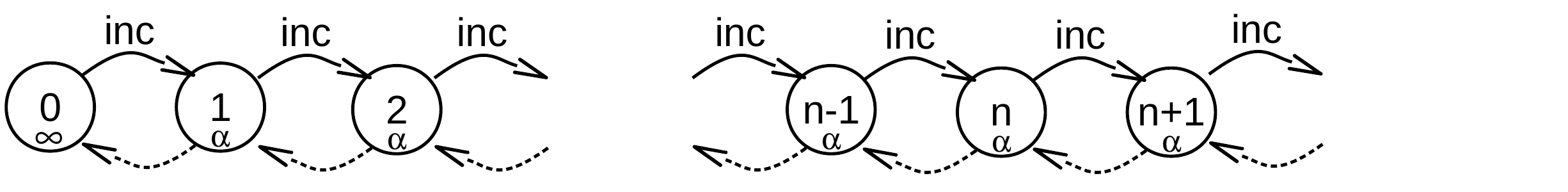}}
		\caption{The state model of a counter.}
		\label{fig:counter}
	}
\end{figure}

Typically, the counter shown in Figure~\ref{fig:counter} may be designed easily using conditional statements. However,  designing this model using patterns has some advantages that we recall at the end of this section.

This counter of numbers  increases by $1$ each time the event $inc$ (increment) occurs. Then, it decreases by $1$ after $\alpha$ u.t from the last event occurred, until the counter reaches the value $0$. The following DEVS model may define such a counter: $\delta_{ext}(): n = n + 1$, $\delta_{int}() : n = n - 1 \text{ if } n > 0$ and $d(): \alpha \text{ if } n > 0 \text{  }\infty \text{ otherwise }$.

At first glance, it is impossible to propose an object oriented design of this counter using the patterns given in this paper. In fact,

the model presents an infinity set of states and transitions as shown in Figure~\ref{fig:counter}. Consequently, it is impossible to design such a model using a finite set of classes without gathering states and transitions that share same properties. We remark that for any state $n>0$ has an internal and external transitions. However the state $n=0$ has only one external transition. Based on that, we propose to design two state classes that represent the counter states $0$ and $n>0$ respectively. Then an external transition class to design the change state from any value to its next value when the event $inc$ occurs ; and two internal transitions classes, each one decreases the counter by $1$ separately when an internal state change occurs. Moreover these classes hold a method boolean condition() allowing to check whether the corresponding transition is fireable or not, defining when the model remains in a state with a positive value or it transients to the state with the null value.

Note that assertions may guarantee that the given design simulates correctly the counter, that we insert in different points of the code.

The power of this object-oriented design consists on dispatching the code of the model in several classes, making easy its readability, reusability and maintainability. For example, it is easy to define a new function that reset the counter to $0$ from any state, or extends its values to negative ones by reusing some classes and by inheritance of others to construct new ones.

\section{Conclusion}
In this work, we addressed a set of designs based on OOP useful for implementing DEVS models. The benefits of such designs are separating of behavior from structure and enhancing the reusability and maintainability of code in order to make easy the correction of bugs and the update of requirements that occur often after the implementation step. The different designs are supported with OCL rules constructed according to DEVS semantics except the legacy of models which needs other techniques like theorem proving or model checking. These rules provide to the programmer a clear and unique interpretation of the given designs. The reusability and maintainability of the simulation based on the discussed designs are shown by carrying out a set of metrics. Recall that the given design patterns implement and interpret correctly the DEVS models. The analysis of simulation traces showed that the different designs give the same behavior.

However, this work has been conducted with compromises due to a lack of some concepts from OOP. For example, there is no way to rewrite the visibility of attributes and methods (modifiers) like declaring the set of submodels of a given model unchanged for DEVS then change it for DS-DEVS. Another inconvenient is the lack of declaring class attributes and methods in a given a class and its subclasses. This fact makes the proposed  event and state event transition patterns less abstract from this view. In addition, the designer should use inheritance to design new concepts of DEVS and its instances which may create confusion. 

This study revealed that the OOP has some lacks to well design DEVS extensions from existing models. Consequently, we believe that a specific oriented language should be defined to DEVS and a special metrics should be identified to measure simulation code according to this language. Then, a formal framework should be designed to support this new language. This, however, doesn't negate the contribution of our work but  will constitute a great challenge and for which we hope that with the DEVS community will explore.

\bibliographystyle{alphaurl}
\bibliography{references}

\end{document}